\documentclass[journal=ancac3,manuscript=article]{achemso}

\usepackage[version=3]{mhchem} 

\author{Thuy An Bui}
\email{thuyan.bui@univie.ac.at}
\author{Gregor T. Leuthner}
\author{Jacob Madsen}
\author{Mohammad R.A. Monazam}
\author{Alexandru I. Chirita}
\author{Andreas Postl}
\author{Clemens Mangler}
\author{Jani Kotakoski}
\author{Toma Susi}
\email{toma.susi@univie.ac.at}
\affiliation[University of Vienna]
{University of Vienna, Faculty of Physics, Boltzmanngasse 5, 1090 Vienna, Austria}

\title[]{Creation of single vacancies in hBN with electron irradiation}

\begin{document}

\begin{abstract}
    Understanding electron irradiation effects is vital not only for reliable characterization of materials using transmission electron microscopy, but increasingly also for the controlled manipulation of two-dimensional materials. Knock-on displacements due to elastic electron backscattering are theoretically straightforward to model, and appear to correctly describe damage in pristine graphene. For semiconducting MoS$_2$, some experimental and theoretical progress has been recently made, indicating that a combination of elastic and inelastic effects appears to be needed to explain experiments. For insulating hexagonal boron nitride (hBN), however, much less is currently known. We measure the displacement cross sections of suspended monolayer hBN using aberration-corrected scanning transmission electron microscopy in near ultra-high vacuum at primary beam energies between 50 and 90~keV. We find damage rates below 80~keV up to three orders of magnitude lower than previously measured at hBN edges under comparatively poorer residual vacuum conditions where chemical etching appears to have been the dominant damage mechanism. Notably, we are now able to show that it is possible to create single vacancies in hBN using electron irradiation, and resolve that boron are almost twice as likely as nitrogen to be ejected below 80~keV. Moreover, any damage at such low energies cannot be explained by pure elastic knock-on, even when accounting for vibrations of the atoms. We thus develop a theoretical description that accounts for a lowering of the displacement threshold energy due to valence ionization resulting from inelastic scattering of probe electrons, and model this using charge-constrained density functional theory molecular dynamics. Although significant reductions in threshold energies are found depending on the constrained charge, quantitative predictions for specific ionization states are currently not possible from first principles. Nonetheless, our findings show the potential for defect-engineering of hBN at the level of single vacancies using electron irradiation.
\end{abstract}

\vskip 2em
Transmission electron microscopy (TEM) is a powerful probe of the atomic structure of materials, and is particularly well suited to directly reveal any defects in the lattice of the specimen. The pioneering work of Crewe~\cite{crewe_scanning_1966} marks the beginning of practical scanning transmission electron microscopy (STEM). From then on, the scanning probe mode was developed as a technique complementary to broad-beam illumination. Due to annular dark-field detection, yielding direct atomic-number contrast, STEM is particularly useful for imaging materials with a mixed elemental composition.

Two-dimensional (2D) materials, a structure family introduced by Novoselov and Geim exfoliating single-layer graphene from graphite~\cite{Novoselov_electric_2004}, provide due to their thinness ideal samples for TEM techniques, especially for atomically resolved quantification of irradiation effects~\cite{susi_quantifying_2019}. This is important because the electron beam can cause changes in the structure of a specimen. Radiation damage in the TEM can be divided into knock-on damage, radiolysis and chemical etching~\cite{egerton_radiation_2004}. The exact mechanism of radiation damage in a specific material is of practical interest as it determines what options are available for minimizing it.

Ballistic displacements (so called knock-on damage) are caused by (quasi-)elastic scattering, where the electron directly transfers momentum to an atomic nucleus. If the resulting kinetic energy of the atom exceeds the displacement threshold energy ({$T_\text{d}$}), the atom is ejected from its lattice position~\cite{banhart_irradiation_1999,kotakoski_electron_2010}. When the knock-on mechanism is predominant, reducing the TEM acceleration voltage will suppress damage~\cite{meyer_accurate_2012}. This process is well understood, especially in graphene~\cite{meyer_accurate_2012, susi_isotope_2016}, although recent studies have suggested that dynamics at its impurity sites cannot be explained by purely elastic effects~\cite{susi_towards_2017, chirita_three-dimensional_2022}.

While radiolysis -- the direct breaking of bonds due to ionization of valence electrons -- is assumed to be predominant in insulators, in conducting materials it is largely quenched due to rapid neutralization. In inorganic solids, both knock-on and ionization effects can take place, sometimes simultaneously. If the damaging effects arise from beam heating or from electrostatic charging of an insulating specimen, reducing the incident beam current can be helpful~\cite{egerton_mechanisms_2012}. To understand the interaction of electrons with different materials, especially semiconductors or insulators where the knock-on mechanism is not the only relevant damage mechanism, a theoretical model describing the process is needed.

For MoS$_2$, recent work has taken the first steps towards quantitative understanding, suggesting excitation-assisted knock-on damage to be the dominant mechanism at electron energies below 80~keV~\cite{zan_control_2013,algara-siller_pristine_2013}. Kretschmer et al. proposed a way to quantify both effects in a combined theoretical framework~\cite{kretschmer_formation_2020}, and the model was further improved by Speckmann et al., removing the ambiguity in model parameters~\cite{speckmann_combined_2023}. Additionally, Yoshimura et al. created a quantum theory of electronic excitation and sputtering and applied it to both MoS$_2$ and hBN~\cite{yoshimura_quantum_2022}. However, the experimental data on hBN they had access to left much to be desired, and thus further developments urgently need accurate quantitative measurements. In particular, the residual vacuum can cause damage to the specimen due to beam-induced chemical etching~\cite{leuthner_scanning_2019}. The electron beam dissociates gas molecules, which can then react with atoms of the lattice. This process is responsible for the growth of pores in graphene~\cite{leuthner_chemistry_2021}, and almost certainly also in hBN -- which is how the only available quantitative data was collected~\cite{cretu_inelastic_2015}.

Defects in hBN are especially interesting for quantum light emission, as its color centers display bright~\cite{tran_quantum_2016} and thermally stable~\cite{kianinia_robust_2017} single-photon emission across a wide spectral range~\cite{tran_robust_2016,bourrellier_bright_2016,gottscholl_initialization_2020}. As such, hBN has potential as a useful platform for quantum computation, information networks and sensors~\cite{deLeon_materials_2021}. Irradiation with lasers~\cite{gan_large-scale_2022}, ions~\cite{choi_engineering_2016}, neutrons~\cite{zhang_discrete_2019} or electrons~\cite{exarhos_optical_2017,su_tuning_2022} have emerged as viable techniques for creating such color centers, and although their exact atomic configuration has been unclear, point defects are certainly involved~\cite{wong_characterization_2015,tran_robust_2016,abdi_color_2018,su_tuning_2022,schauffert_characteristics_2023}. However, considering the rapid damage and creation of pores under typical TEM conditions~\cite{kotakoski_electron_2010,cretu_inelastic_2015}, it has not been clear whether single vacancies can be controllably created.

\section{Results and Discussion}
We carried out measurements in ultra-high vacuum (UHV), which suppresses unwanted chemical etching and allows us to obtain separate cross sections for individual B and N at atomic resolution. Chemical etching occurs already at a typical residual vacuum pressure of $\sim$10$^{-7}$ mbar~\cite{leuthner_scanning_2019}, which can even change the dominant termination of pore edges in electron-irradiated graphene~\cite{leuthner_chemistry_2021}. Indeed, the cross sections we measure are up to three orders of magnitude lower than what was reported for atoms at pore edges~\cite{kotakoski_electron_2010,cretu_inelastic_2015}, suggesting previous experiments have been dominated by chemical etching and showing that single vacancies can indeed be created by electron irradiation at intermediate energies.

\begin{figure}
    \centering
    \includegraphics[width=1\textwidth]{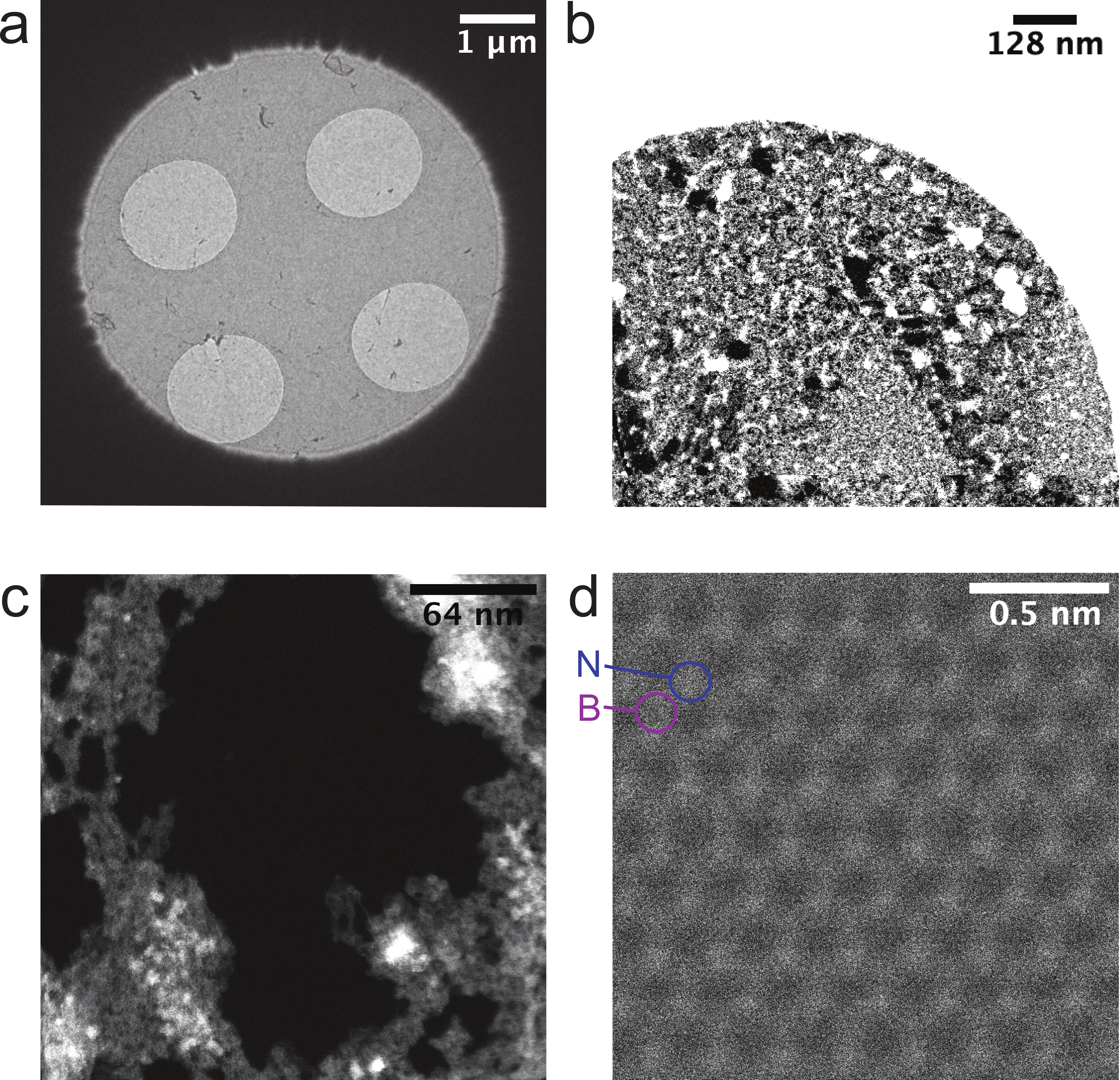}
    \caption{\textbf{STEM images showing an hBN sample at different magnifications.} \textbf{(a)} Bright-field Ronchigram image of the sample support foil covered with monolayer hBN seen within the virtual objective aparture. \textbf{(b)} MAADF image of single layer of hBN suspended over a hole. The darkest area corresponds to a clean lattice, whereas the brighter areas are covered with surface contamination. The white area surrounding the sample corresponds to the support foil. \textbf{(c)} Area of clean monolayer hBN. \textbf{(d)} Atomic resolution MAADF image of monolayer hBN.}
    \label{fig:hBN_overview}
\end{figure}

\begin{figure}
    \centering
    \includegraphics[width=12cm]{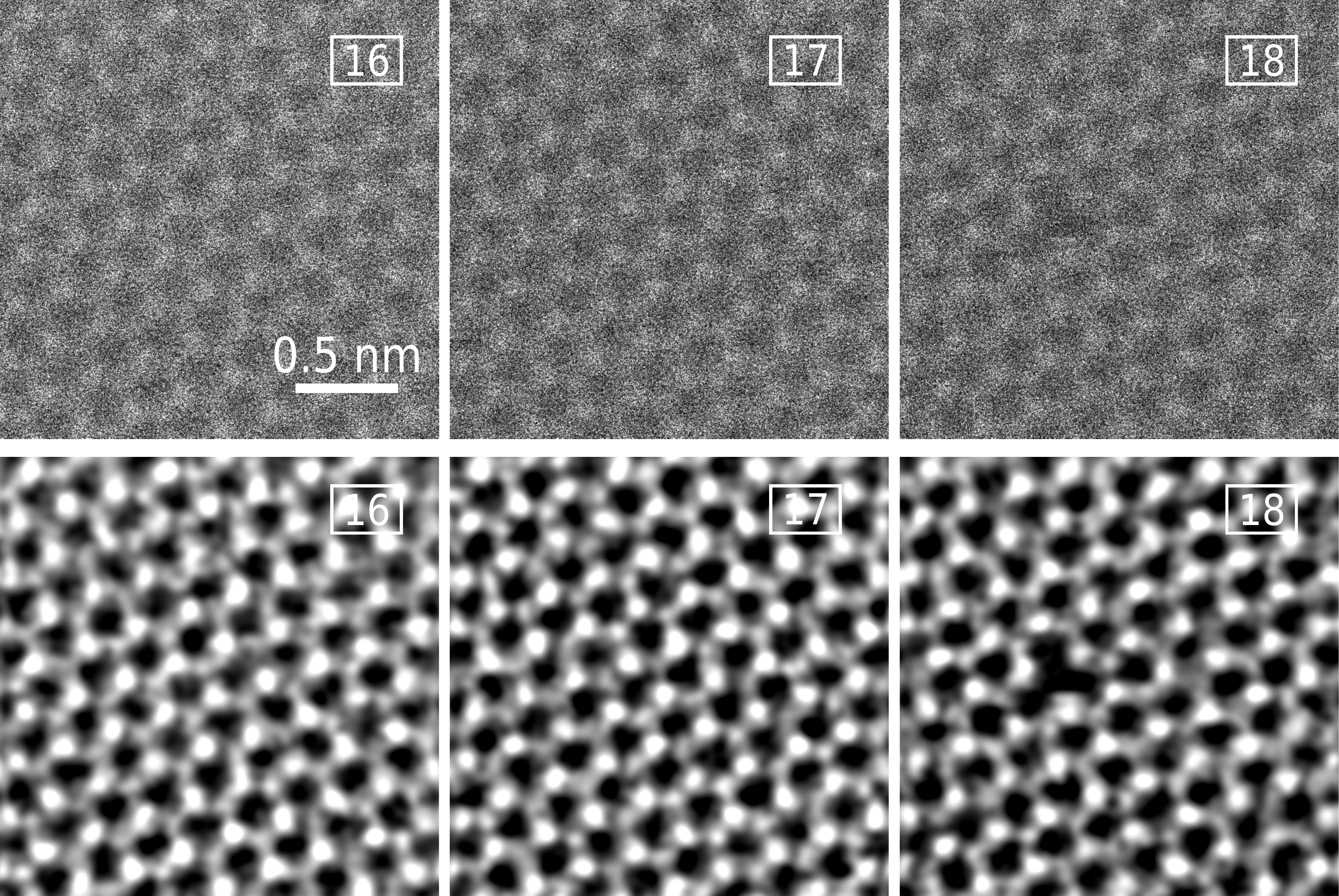}
    \caption{\textbf{STEM MAADF images showing the creation of a single B vacancy.} The first two unprocessed images (upper row) are from the image sequence before the defect appears. The overlaid numbers are the number of each image in the series. Image 18 shows the first B vacancy. Images in the lower row have been processed using the double-Gaussian filter ($\sigma_1 = 0.23, \sigma_2 = 0.15, \text{weight} = 0.28$).}
    \label{fig:defect_B}
\end{figure}

\begin{figure}
    \centering
    \includegraphics[width=12cm]{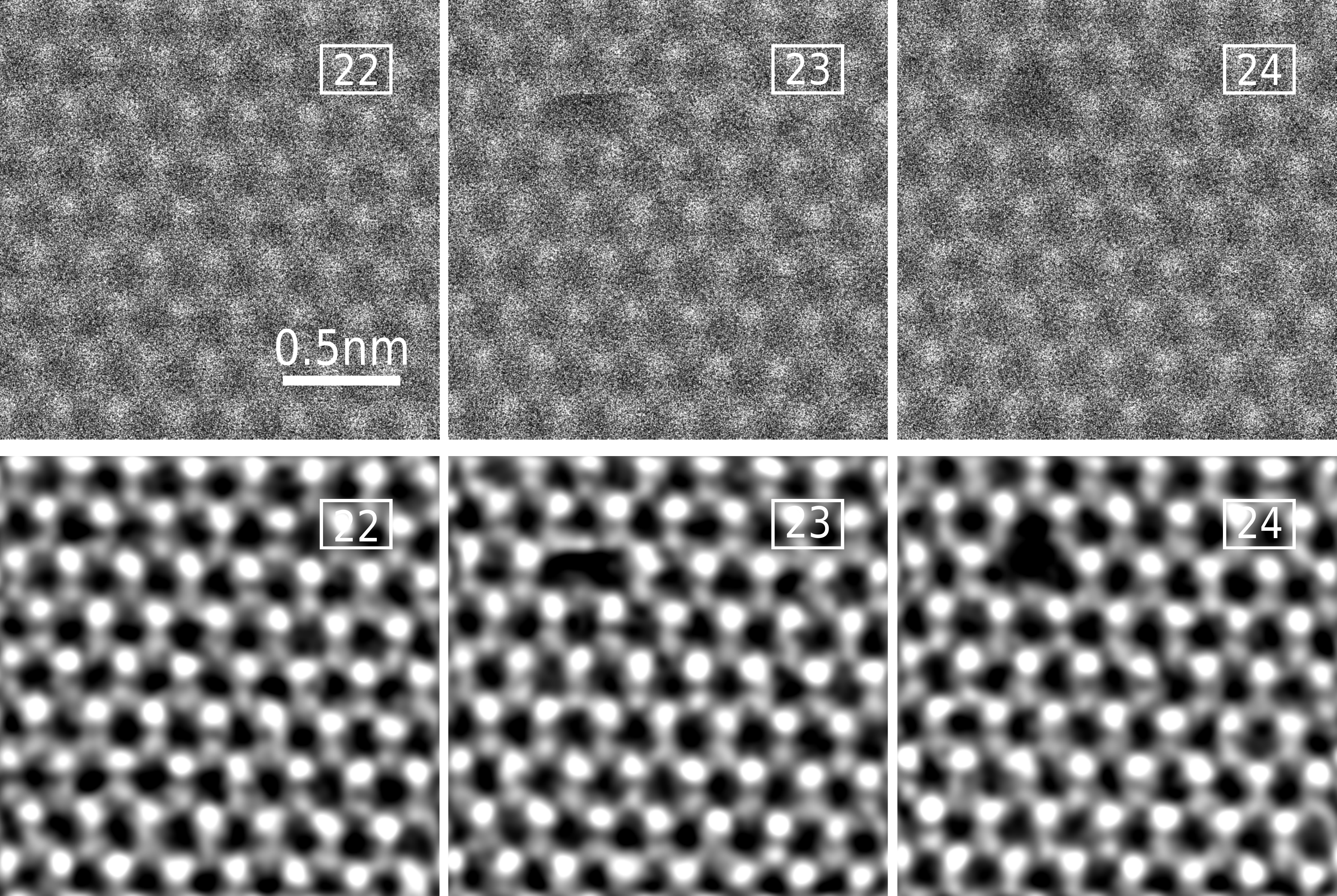}
    \caption{\textbf{STEM MAADF images showing the creation of a single N vacancy.} First unprocessed image (upper row) is from the image sequence before the defect appears. The overlaid numbers are the number of each image in the series. Image 23 shows where N atom ejection occurs during scanning, and image 24 shows the first N vacancy. Images in the lower row have been processed using the double Gaussian filter ($\sigma_1 = 0.24, \sigma_2 = 0.16, \text{weight} = 0.29$).}
    \label{fig:defect_N}
\end{figure}

Samples were prepared using commercial hBN grown by chemical vapor deposition following the procedure of Ref.~\citenum{ahmadpour_monazam_substitutional_2019} (see Methods), containing clean areas up to 100 nm in size. An overview at different length scales is provided in Fig.~\ref{fig:hBN_overview}. Exemplary STEM medium-angle annular dark-field image series recorded at an energy of 55~keV are displayed in Figs.~\ref{fig:defect_B} and \ref{fig:defect_N}, which show both raw data as well as double-Gaussian filtered images~\cite{krivanek_atom-by-atom_2010}. Each frame has 512$\times$512 pixels for a field of view of typically 2$\times$2~nm$^2$ and a scanning time of roughly 0.26~s. While scanning even faster would be beneficial, we found that this would have degraded the signal to noise ratio too much for reliable analysis.

Only cases where we could unambiguously identify the first ejected atom were included in the analysis. Cases where two separated vacancies of the same element were present in the first defected frame were counted as only a single vacancy, since the creation of the first could have facilitated the second. Just twice in all our data were two separated N and B atoms missing in the first defective frame, illustrating a high probability to create single vacancies.

To identify the number of frames without a defect and the position where the first vacancy occurred, a machine-learning algorithm based on a convolutional neural network was used to assist human analysis (see Methods and Supporting Information). The experimental displacement cross section ~\cite{banhart_irradiation_1999,zobelli_electron_2007, meyer_accurate_2012, susi_quantifying_2019} can be derived from the electron dose $\phi$, which consist of dose rate $\phi_R$ multiplied by time until the defect $t_d$ (detailed calculation can be found in the Supporting Information). It corresponds, following Poisson statistics, to the expectation value of the dose for the distribution arising from a large number of repeated measurements.

Dose rates were evaluated by estimating the beam current impinging on the sample via continuous measurement of the current falling on the virtual objective aperture of the microscope, which is saved with the image metadata. This current in turn was calibrated against the beam current measured at the drift tube of the electron energy-loss spectrometer. While typically the beam current is measured only occasionally, we recently found this may lead to significant errors in the estimated doses, and thus have used here the statistical measurements described in Ref.~\citenum{speckmann_combined_2023}.

The displacement cross section $\sigma$ describes the probability of an energetic electron scattering from an atomically thin sample to lead to a displacement of one of its atoms, calculated as the inverse of the Poisson expectation value multiplied by the material-specific areal density $\rho_{\text{hBN}}$ (see Supporting Information for more detail). To obtain accurate values of such a stochastic process, measurements of pristine areas need to be repeated until sufficient statistics are collected. Our analysis includes 664 valid series for N and 337 for B (Supporting Fig.~\ref{sifig:expfit} displays histograms of the data).

\subsection{Simulation of ionization-assisted displacement thresholds}

To study the role that valence ionization might play in the knock-on damage process, we performed density functional theory molecular dynamics (DFT/MD) simulations of the ejection of either N or B atoms from hBN following our established methodology~\cite{susi_atomistic_2012,susi_isotope_2016} (see Methods for more detail). In addition to ground-state calculations previously reported for hBN~\cite{kotakoski_electron_2010}, we now use constrained DFT (cDFT) to confine a point charge on the ejecting atom to quantify its effect on the displacement threshold energy. We note that this description is not entirely realistic, as the confining potential has no direct physical interpretation, and it is enforced for the full duration of the simulation (i.e. there is no charge neutralization). Accurately describing dynamics with a localized valence hole in periodic supercell is a difficult challenge, and thus this approach should be considered as a pragmatic if imperfect approximation.

In earlier work, Kotakoski et al.~\cite{kotakoski_electron_2010} calculated the displacement thresholds for B and N atoms in the pristine structure and at different vacancy edges in various total charge states. For pristine structures in the ground state, they obtained $T_{\text{gs}}^B = 19.36$~eV and $T_{\text{gs}}^N = 23.06$~eV, somewhat different than our values of $T_{\text{gs}}^B = 20.15$~eV and $T_{\text{gs}}^N = 22.25$~eV. We recently noticed that the specific value of Fermi smearing used in GPAW seems to have an influence of about this magnitude on the absolute threshold energy values~\cite{chirita_three-dimensional_2022}, though also the difference between the two elements seems also to be slightly different (3.7~eV vs. 2.1~eV). Nonetheless, the two calculations are in general agreement.

\begin{figure}[t]
    \centering
    \includegraphics[width=12cm]{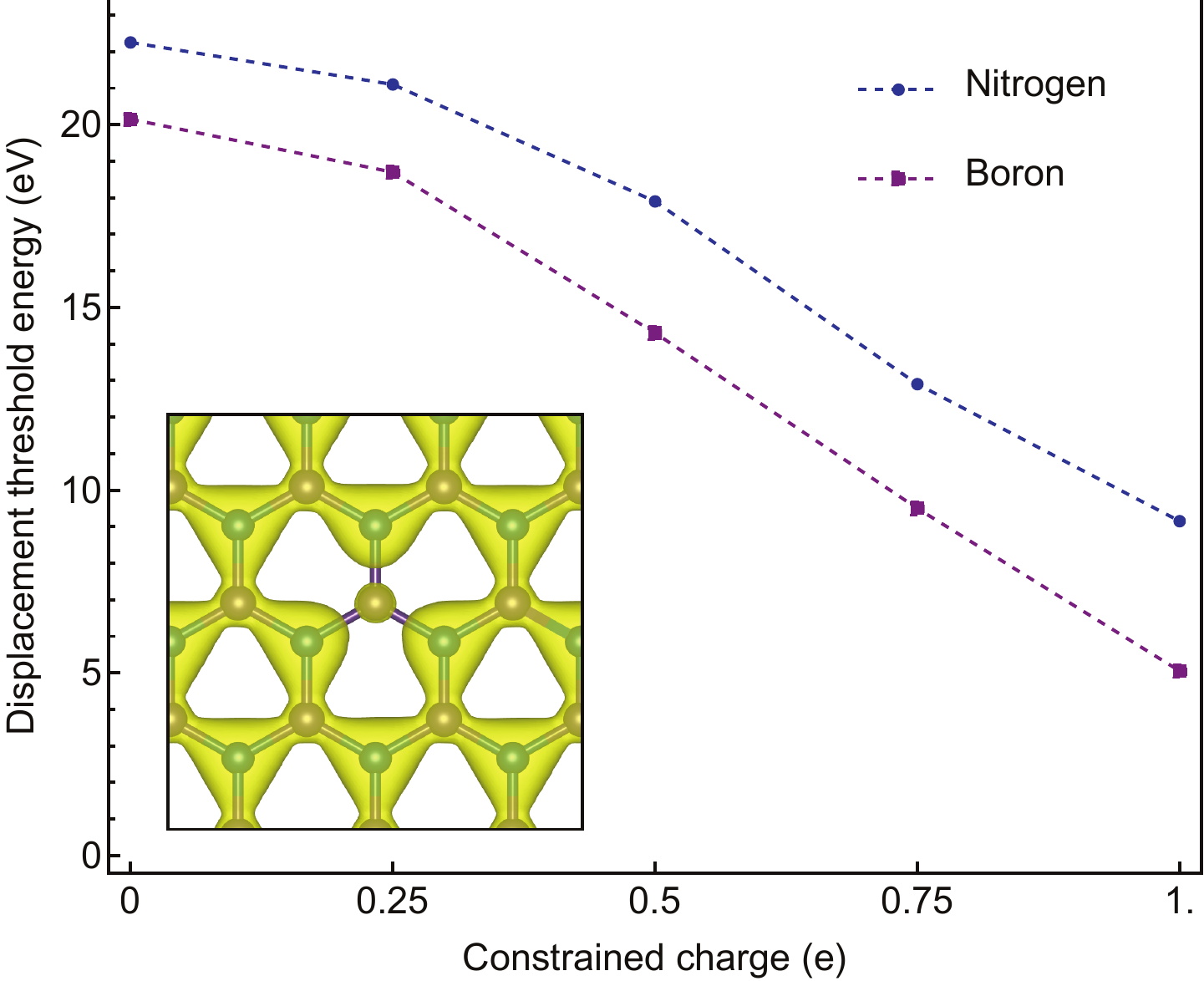}
    \caption{\textbf{Reduction of the displacement threshold energy $T_\mathrm{d}$ as a function of charge constrained on the ejecting element.} The inset shows a charge-density isosurface of a 1.0 hole constrained on the B atom at the start of a cDFT/MD simulation (isovalue 0.15 e/\AA$^3$).}
    \label{fig:cDFT_thresholds}
\end{figure}

For a full positive constrained elementary charge (corresponding to a missing electron), we find that the displacement threshold for N drops from 22.25~eV to 9.15~eV ($-$59\%), while that for B drops from 20.15~eV to 5.05~eV ($-$75\%). Since it is likely that even in an insulating material such as hBN, some part of the ionized charge is neutralized during the roughly 100~fs it takes for the impacted atom to fully displace, we also considered partial charges to study the systematic behavior of the threshold energy. These results are shown in Fig.~\ref{fig:cDFT_thresholds}: after a slighter reduction for the 0.25 hole, the values drop roughly linearly with a slope of -16.3~eV/e for N and -18.2~eV/e for B.

\subsection{Theoretical cross section with ionization}

From the displacement threshold energy $T_\text{d}^{\text{gs}}$ one can estimate the displacement cross section $\sigma$ under known electron-beam conditions using the McKinley-Feshbach formalism \cite{mckinley_coulomb_1948} for Coulombic scattering of relativistic electrons \cite{su_engineering_2019}. The corresponding total elastic displacement cross section~\cite{dugdale_some_1954,zobelli_electron_2007}, with the velocities $v$ of the atoms accounted for using phonon modeling~\cite{susi_isotope_2016} (as described in the Methods), can be written as
\begin{equation} 
\begin{split}
    \sigma_{\text{KO}} (E_e, v) = 4 \pi \left( \frac{Z e^2}{4 \pi \epsilon_0 2 \gamma m_0 c^2 \beta^2 } \right)^2
    \bigg\{ \left(\frac{E_{\text{max}}(v)}{T_\text{d}^{\text{gs}}} -1\right) - \beta^2 \ln{\left(\frac{E_{\text{max}}(v)}{T_\text{d}^{\text{gs}}} \right)} \\
    + \pi Z \alpha \beta \left[2 \left( \sqrt{\frac{E_{\text{max}}(v)}{T_\text{d}^{\text{gs}}}} -1 \right) - \ln{\left( \frac{E_{\text{max}}(v)}{T_\text{d}^{\text{gs}}}\right)} \right]  \bigg\},
\end{split}
    \label{eq:sigma_ko_elastic}
\end{equation}
where $E_\text{max}(v)$ is the maximum transmitted energy at out-of-plane velocity $v$, $Z$ is the atomic number, $e$ the elementary charge, $\epsilon_0$ the electric constant in vacuum, $m_0$ the electron rest mass, and $\alpha$ the fine-structure constant. The relativistic Lorentz factor is $\gamma = \sqrt{1 - \beta^2}$, and $\beta$ depends on the primary beam energy $E_e$ as
\begin{equation}
    \beta (E_e) = \sqrt{1 - \left(1 + \frac{E_e}{m_0 c^2} \right)^{-2}}.
    \label{eq:beta}
\end{equation}

Previously only the ground-state dynamics were considered. However, beam-induced excitation may reduce the bonding energy of the atoms in the lattice, thus reducing the threshold energy. While the displacement caused by an elastic scattering process (also known as a knock-on displacement) is well understood and theoretically described, the role of inelastic electron-electron scattering still remains unclear. A classical theoretical model was derived by Bethe \cite{bethe_zur_1930}, later on complemented relativistically by M\o ller \cite{moller_zur_1932}. A modification of Bethe's theoretical formula by using impact parameter and substitutions to fit experiments was done by Williams~\cite{williams_applications_1933}, which is the source of the inelastic scattering cross section reprinted in Williams \& Carter~\cite{williams_transmission_2009}. However, despite being well known, that formula was provided in CGS units and the definition of the included velocity term was not explicitly given. It is therefore worth reprinting it here, with some modification.

In SI units, the ionization cross section for shell $s$ is (see Supporting Information)
\begin{equation}
    \sigma_{\text{Bethe}}(E_e, s, i) = \left( \frac{2 \pi e^4 b_s n_s}{(4 \pi \epsilon_0)^2 m_0 c^2 \beta^2 E^{\text{ion}}_i} \right) \bigg\{ \ln{ \left[c_s \left( \frac{m_0 c^2 \beta^2}{2 E^{\text{ion}}_i} \right) \right]} - \ln{\left(1 - \beta^2 \right)} - \beta^2 \bigg\},
    \label{eq:sigma_bethe}
\end{equation}
where $n_s$ is the number of electrons in the subshell that is ionized, $b_s$ and $c_s$ are a priori unknown constants for that shell, and $E^\text{ion}_i$ is in our treatment the $i$th ionization energy. For a given electron shell and known ionization energies, measured atomic cross section data can be fitted to obtain the constants. As this data is available for low primary beam energies (up to 1~keV for B~\cite{kim_ionization_2001} and 5~keV for N\cite{kim_ionization_2002}), care must be made to fit the highest-energy data points as well as possible so that the slope extending to our experimental energy range is correct (see Supporting Information for detail).

As will be discussed in the next section, it turns out that to be able to fit experimental hBN data, we needed to include two ionized states for boron (Supporting Fig.~\ref{sifig_boron}) and three for nitrogen (Supporting Fig.~\ref{sifig_nitrogen}). The calculated ionization cross sections, extrapolated to our experimentally relevant primary beam energies, are shown in Supporting Fig.~\ref{sifig:ionization_B_N}. The cross section values are on the order of $10^{6}$~barn and show the characteristic $1/E$ dependence, though it should be noted that their ratio remains almost constant at  0.68--0.69 across the TEM-relevant range of energies (20--100~keV). However, as we will see when trying to describe the experimental displacement data for hBN, we had to modify the probabilities of each ionized state to obtain a good match.

\subsection{Fitting experimental cross sections}

\begin{figure}
    \centering
    \includegraphics[width=0.93\textwidth]{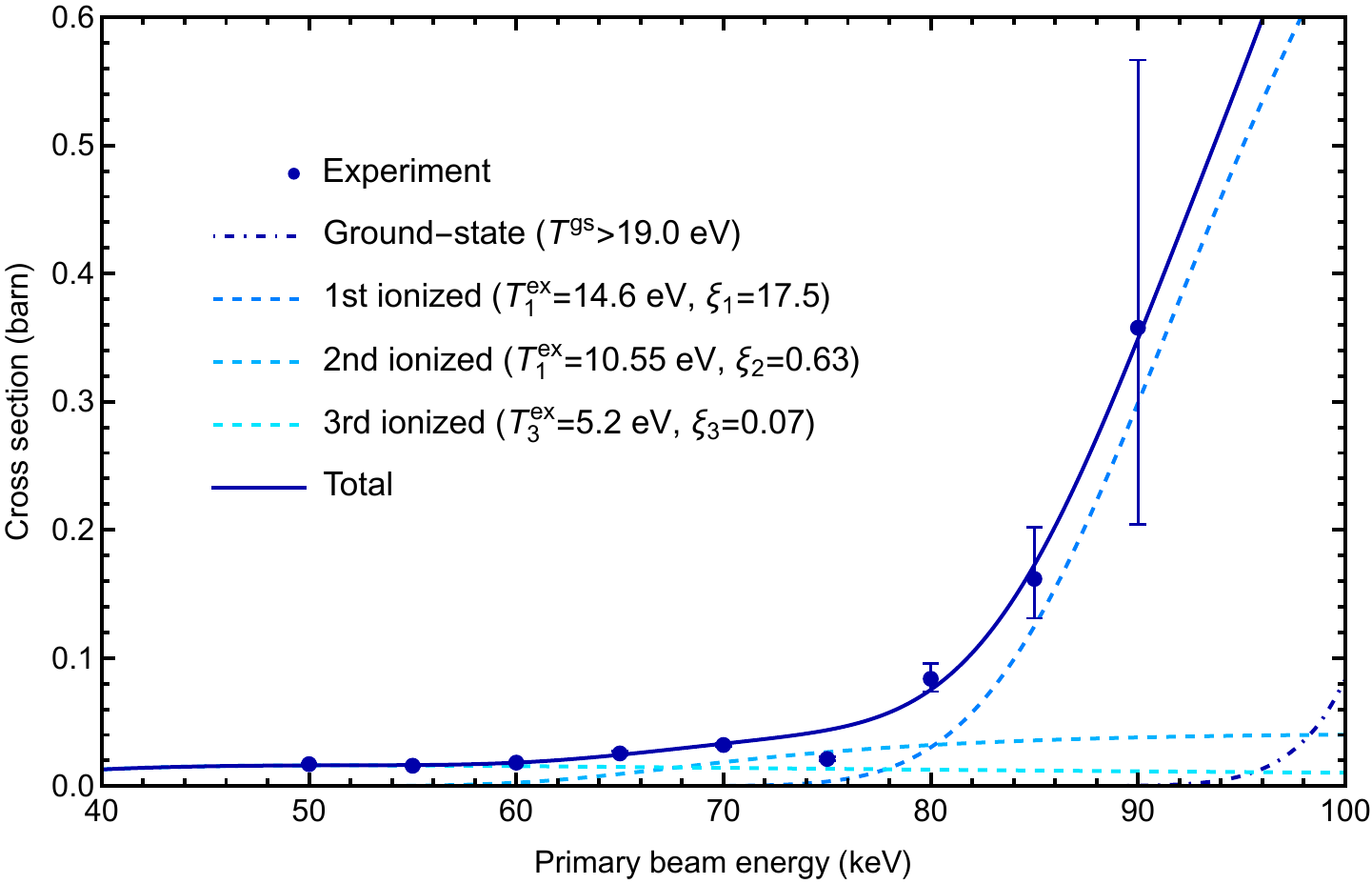}
    \caption{\textbf{Displacement cross sections for nitrogen.} Theoretical fits to experimentally measured datapoints for nitrogen.}
    \label{fig:fit_N}
\end{figure}

In hBN the mass difference between the two constituent elements is more than 20\%. For N, we only need to consider the stable $^{14}$N isotope. However, boron has two abundant stable isotopes with a significant relative mass difference, and thus we need to consider separately $^{10}$B and $^{11}$B with ca. 20 atom-\% and 80 atom-\% where the masses are 10.01 u and 11.01 u. Due to the greater mass of nitrogen, elastic energy transfers from primary beam electrons at our experimental energies are suppressed at displacement threshold energies below 19~eV, significantly below what DFT predict (Fig.~\ref{fig:fit_N}), which means that the displacement of nitrogen from the ground state cannot happen at energies up to 90~keV.

\begin{figure}
    \centering
    \includegraphics[width=0.93\textwidth]{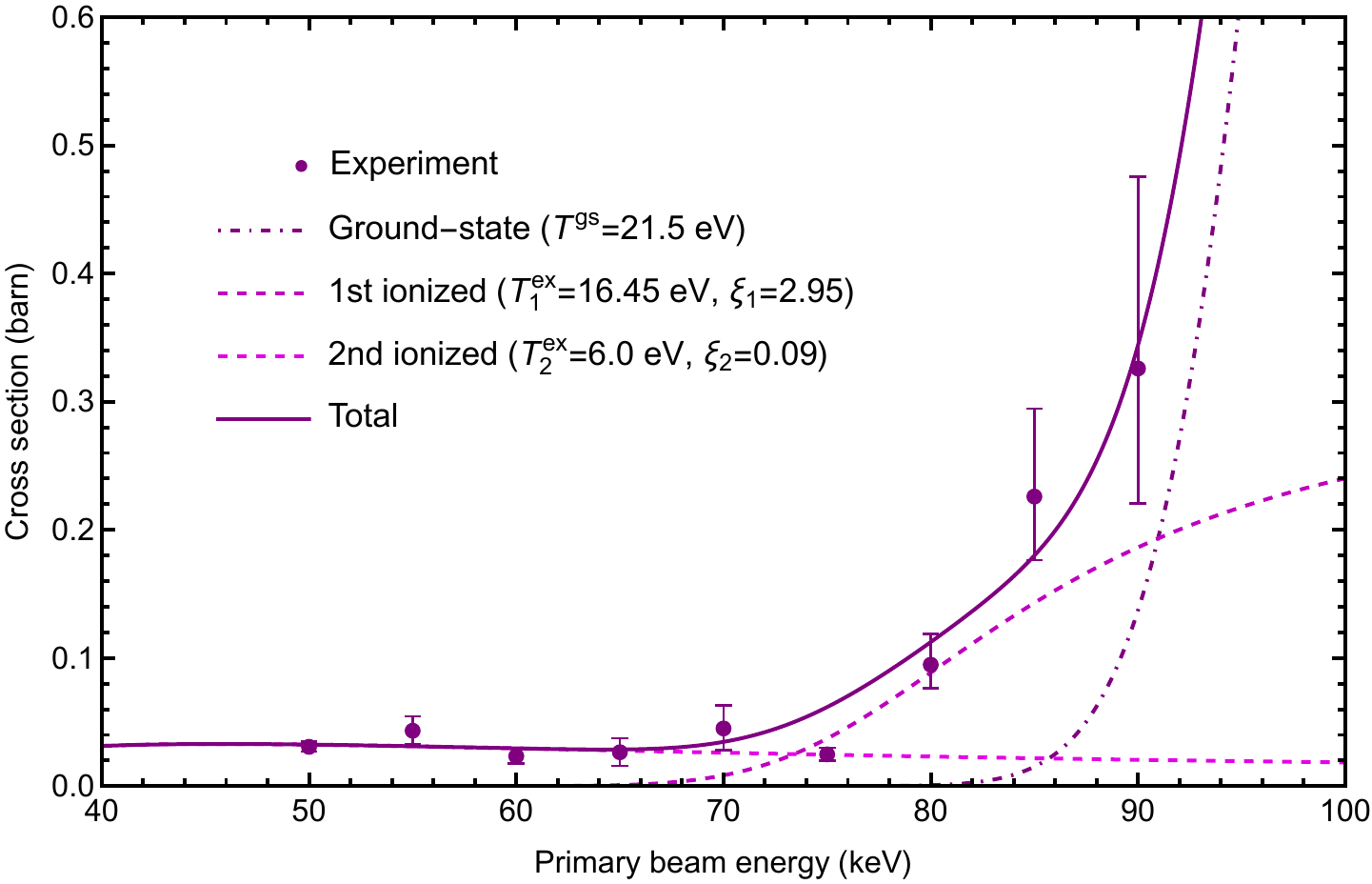}
    \caption{\textbf{Displacement cross sections for boron.} Theoretical fits to experimentally measured datapoints for boron.}
    \label{fig:fit_B}
\end{figure}

Figures~\ref{fig:fit_B} and \ref{fig:fit_N} show a fitting of the experimental data with our theoretical model that combines the earlier mentioned elastic and inelastic scattering processes. (The experimental points at 75~keV for both elements were excluded from the fits as outliers, and we suspect that the experimental current at that energy does not match the measured calibrations.) The total displacement cross section $\sigma_{\text{tot}} $ is calculated as a additive combination of the ground $\sigma_{\text{KO}} (E_e,v,T_\text{d}^{\text{gs}})$ and excited $\sigma_{\text{KO}} (E_e,v,T_{\text{d},i}^{\text{ex}})$ states multiplied by the corresponding Bethe cross sections and the areal density $\rho_{\text{at}}$
\begin{equation}
    \sigma_{\text{tot}} = \sigma_{\text{KO}} (E_e, v, T_\text{d}^\text{gs}) + \sum_{i=i_1}^{i_\text{max}} \sigma_{\text{KO}} (E_e, v, T_{\text{d},i}^\text{ex}) \cdot (\sigma_{\text{Bethe}}(E_e, s, i) \cdot \xi_i) \cdot \rho_{\text{at}},
    \label{eq:sigma_tot}
\end{equation}
where $i$ indexes the ionization state (up to $i_\text{max}$ = 2 for B and 3 for N) and $\xi_i$ is a parameter accounting for inaccuracies in the ionization cross sections as well as finite lifetimes of the excited states. As the material consists of two elements, two separate analyses were conducted, with the above-mentioned isotope-weighted average used for boron.

It is clear from the plots that the theoretically predicted ground-state threshold energies cannot alone describe the observed cross section values -- indeed, for nitrogen no damage would be expected up to 90~keV even if the ground-state threshold energy was as low as 19~eV -- whereas the combined model does provide a good fit with the data. However, we did have to significantly increase the atomic ionization cross sections for the first ionized state: 2.92 times for boron and as much as 18.4 times for nitrogen. Only values smaller than 1 can be explained by a finite lifetime of the excitation~\cite{yoshimura_quantum_2022,speckmann_combined_2023}.

\subsection{Discussion}

We must note that the model described above is simplified in two important ways. First, the ionization energies used for the Bethe cross section in Eq.~\ref{eq:sigma_bethe} are tabulated for sequential ionization, ie. the removal of further electrons from an already ionized atom, which is unlikely to be the case experimentally for hBN (and would require the multiplication of ionization probabilities). We consider the model as an approximate description of the more likely experimental situation, namely multiple ionization by the same electron. Second, the probability of being in an ionized state should be subtracted from the ground-state process, as was done in Ref.~\citenum{speckmann_combined_2023}. However, in our case this is not possible, as allowing $\xi_1$ to take values above 1 -- indicating that the fitted atomic ionization cross sections underestimate the probability of ionization in hBN or that the process cannot correctly be described using this model -- would mean that probabilities do not sum correctly. We have therefore neglected this correction, which would slightly lower the fitted ground-state threshold energies.

Using ionization cross-section data measured for atoms is clearly not ideal for describing a periodic solid with significant ionic character. Measurements on more closely related molecules such as borazine, or even better, hexaborane, might provide an improved reference model. Even better would be direct measurements on hBN itself at the experimentally relevant TEM energies, but this is clearly not feasible using crossed beams of electrons and fast atoms or molecules which are the standard method to measure impact ionization cross sections~\cite{brook_measurements_1978}. Nonetheless, due to the relatively weak dependence of the extrapolated high-energy tail of the cross sections on the details of the fitting of the low-energy data, we agree with Kretschmer et al.~\cite{kretschmer_formation_2020} that this approach can provide a useful approximation.

The range of fitted excited-state thresholds falls within the simulated cDFT/MD values, but it is difficult to predict them for specific ionization states from first principles. Constrained DFT as a starting point for Ehrenfest dynamics is not feasible due to the artificial nature of the constraining potential: time-dependent dynamics without the constraint lead to rapid neutralization of the charge on a timescale much shorter than the time to displace an atom. Thus, further theory development based on the precise cross section values reported here will be needed for a correct first-principles description of hBN irradiation damage.

A recent study has suggested that nitrogen vacancies are the active color center in hBN~\cite{su_tuning_2022}. Our precise experimental measurements seem to indicate that it is quite challenging to exclusively create N vacancies, as the cross section for knocking out B atoms is higher across the range of studied energies. Nonetheless, even parallel-beam irradiation at 200~keV can apparently controllably create optically active regions in hBN~\cite{su_tuning_2022}. This may be explainable by the extrapolation of our cross sections to higher primary beam energies: at 200~keV, due to its lower threshold energies and the decreased importance of their mass difference, the ejection of N becomes more likely than that of B (Supporting Fig.~\ref{sifig:higherE}). This could result in a greater number of N vacancies at a fixed irradiation dose. Additional data at primary beam energies higher than 90~keV would be helpful to further reduce the related uncertainties, but the rapidly increasing damage cross section means that even at 90~keV, in many data series larger defects appear already as the very first defect, making it impossible to assign the element to the first ejection and thus to obtain reliable estimates.

Now that it is clear that hBN is remarkably radiation resistant at primary beam energies below 80~keV, the creation of color centers could potentially be greatly optimized by site-selective irradiation using a focused STEM electron probe~\cite{susi_manipulating_2017} coupled with low-dose imaging and automation enabled by machine learning~\cite{roccapriore_probing_2022}.

\section{Conclusion}

Despite its insulating nature, monolayer hexagonal boron nitride is surprisingly stable under electron irradiation when chemical etching can be prevented. Single boron and nitrogen vacancies can be created at intermediate electron energies, although boron are twice as likely to be ejected due to its lower mass. However, we predict that at energies of 200~keV and above, nitrogen in turn becomes easier to eject. Providing reliable experimental measurements to develop theoretical models can lead to a better understanding of ionization damage in non-conducting materials more broadly, where a combination of inelastic and elastic scattering appears to be active. Further, point defects in hBN are of great current interest due to their single-photon emission properties, and it may be possible to use electron irradiation to purposefully create them. New opportunities for atomically precise manipulation, until now demonstrated for dopant atoms in graphene and in bulk silicon, may also be uncovered.

\section{Methods}
\subsection{Sample preparation}
Commercial monolayer hBN synthesized with chemical vapor deposition~\cite{kim_synthesis_2012} on copper foil purchased from Graphene Supermarket (CVD hBN on copper foil) was transferred onto a conventional Au Quantifoil TEM grid (R 1.2/1.3) (Supporting Figure~\ref{sifig:smple_prep}a). To attach the TEM grids onto hBN, a drop of isopropanol (IPA) was put on top of the already placed grid (Supporting Figure~\ref{sifig:smple_prep}b). For a successful transfer, uniform adhesion between the substrate and the TEM grids is necessary. To etch away the copper foil, we used an 10\% iron(III)-chloride(FeCl$_3$) acid solution (Supporting Figure~\ref{sifig:smple_prep}c). The sample was introduced to vacuum and baked overnight at ca. 150\textdegree C before measurements were conducted. Between measurements, the sample was stored in the CANVAS ultra-high vacuum system~\cite{mangler_materials_2022}.

\subsection{Scanning transmission electron microscopy and analysis}
STEM measurements were performed with an aberration-corrected Nion UltraSTEM100 at primary beam energies ranging from 50 to 90 keV. Image series were recorded using a medium-angle annular dark field (MAADF) detector with a collection angle of 80--200 mrad. The base pressure at the sample was below 10$^{-9}$ mbar, suppressing any chemical etching~\cite{leuthner_scanning_2019}. 

A machine-learning algorithm based on a convolutional neural network trained using simulated images~\cite{trentino_atomic-level_2021} was used to detect the position of the first occurring defect in each recorded image series. The intensity difference between the N and B sublattices was also used to determine which element was ejected, and the analysis results were verified by manual inspection (see Supporting Information for detail).

\subsection{Density functional theory molecular dynamics}
To perform the simulations, we used the Atomic Simulation Environment~\cite{larsen_atomic_2017} for Velocity-Verlet dynamics with a timestep of 0.3\,fs on a 7$\times$7$\times$1 graphene supercell, with forces from a GPAW~\cite{enkovaara_electronic_2010} calculator with the PBE functional~\cite{perdew_generalized_1996}, a localized \textit{dzp} basis set, a 5$\times$5$\times$1 Monkhorst-Pack $\mathbf{k}$-point grid, and a Fermi-Dirac smearing of 0.025\,eV. 

The GPAW implementation of cDFT imposes a localized constraining potential via an atom-centered Gaussian function~\cite{melander_implementation_2016}, whose weight is iteratively optimized to achieve the desired charge on the atom as assessed by Hirshfeld partitioning~\cite{hirshfeld_bonded-atom_1977}. For the ground state, we considered both spin-paired and spin-polarized calculations, whereas the charge-constraint was always imposed on one spin channel. We also simulated 6$\times$6$\times$1 and 5$\times$5$\times$1 supercells, and confirmed that the calculated threshold energies are converged within 0.1~eV.

\subsection{Phonon density of states}
Phonon calculations were performed using density functional perturbation theory (DFPT)~\cite{baroni_phonons_2001, gonze_dynamical_1997} as implemented in the ABINIT code~\cite{gonze_abinit_2009}. The lattice structure was first optimized down to an energy difference of 10$^{-10}$ Ha with a \textbf{k}-point mesh of 20$\times$20$\times$1 and an energy cut-off of 55 Ha. Exchange and correlation were described using the local density approximation and Troullier–Martins norm-conserving pseudo-potential~\cite{troullier_efficient_1991}. 

For calculating the Hessian and dynamic matrix, the ground-state wave functions were converged to within 10$^{-18}$ Ha$^2$ with a \textbf{k}-point mesh of 40$\times$40$\times$1. Phonon density of states (DOS) was interpolated using the Gaussian method with a smearing of 6.5$\times$10$^{-5}$ Ha, and assigned to the B and N sublattice to estimate their out-of-plane mean-square velocities as described in Ref.~\citenum{susi_isotope_2016}. The calculated phonon DOS is shown as Supporting Fig.~\ref{sifig:phononDOS}.

\section{Supporting Information}
Supporting Information: Schematic illustration of sample preparation, detailed description of experimental cross section, histograms and numerical values of experimental data, description of machine-learning algorithm, detailed description of theoretical ionization cross section, fitting of atomic impact ionization data, extrapolated theoretical valence ionization cross sections for B and N, extrapolated total displacement cross sections for hBN, and plot of the phonon density of states (PDF; numerical data included also as comma-separated files). Calculations and fitting of the theoretical cross section models to the data (Wolfram Mathematica 13.1 notebook) can be accessed on the Wolfram Notebook Archive at \url{https://notebookarchive.org/2023-05-4lw178w}. The raw experimental STEM datasets can be accessed on the University of Vienna institutional repository Phairda online at \url{https://doi.org/10.25365/phaidra.399}.

\begin{acknowledgement}
T.A.B, J.M., M.R.A.M., A.I.C., A.P., and T.S. were supported by the European Research Council (ERC) under the European Union’s Horizon 2020 research and innovation programme (Grant agreement No.~756277-ATMEN). We gratefully acknowledge computational resources provided by the Vienna Scientific Cluster (VSC). 
\end{acknowledgement}

\newpage
\renewcommand{\thefigure}{S\arabic{figure}}
\setcounter{figure}{0}

\renewcommand{\thetable}{S\arabic{table}}
\setcounter{table}{0}

\begin{suppinfo}

\subsection{Sample preparation}

\begin{figure}
    \centering
    \includegraphics[width=1\textwidth]{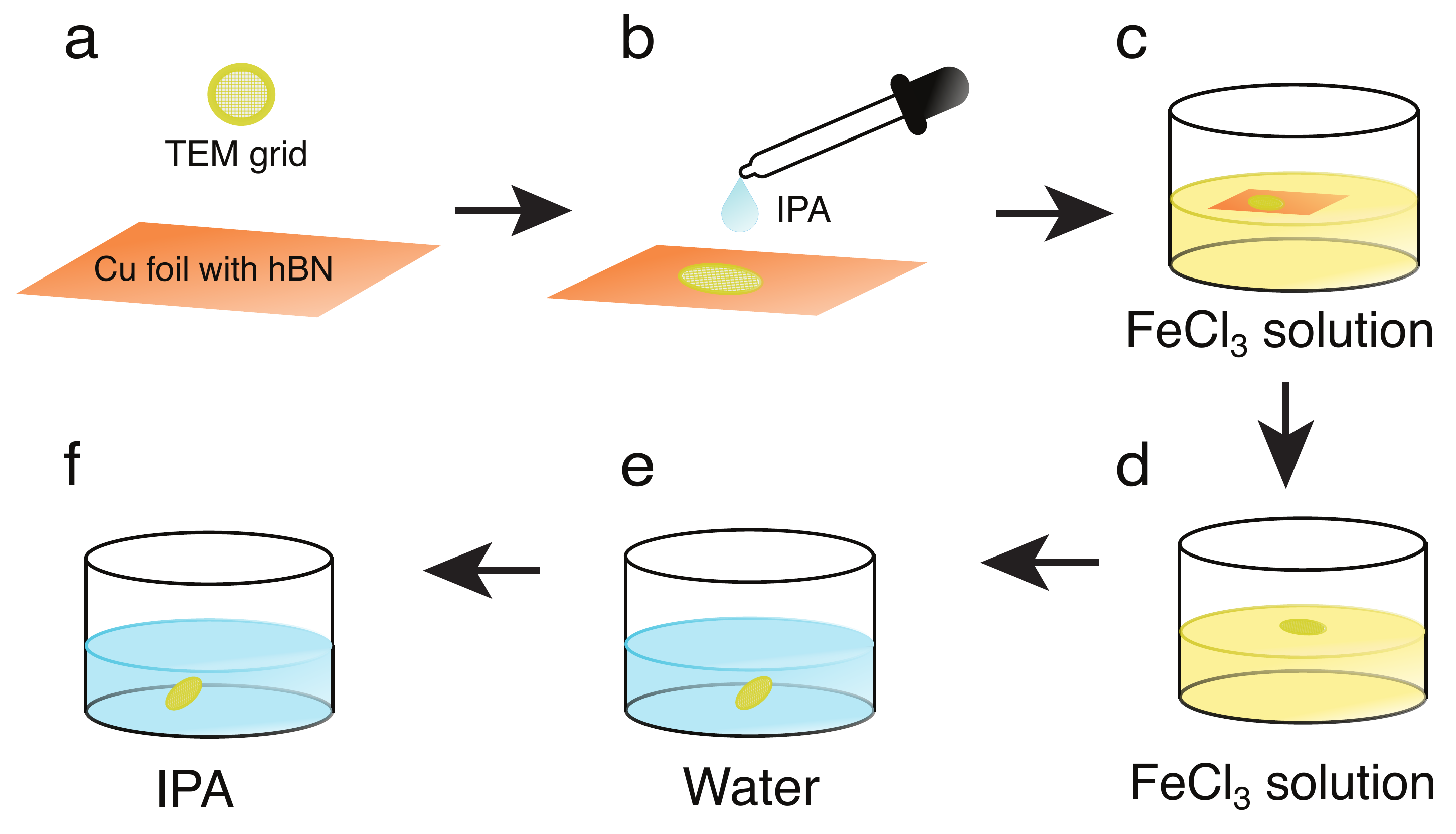}
    \caption{\textbf{Sample preparation.} \textbf{(a)} Gold TEM grid and CVD-grown hBN on Cu foil. \textbf{(b)} Attaching the TEM grid onto the Cu foil with isopropanol (IPA). \textbf{(c)} TEM grid on top of the Cu foil on the surface of a FeCl$_3$ solution. \textbf{(d)} Solution with the sample (TEM grid) after some days. \textbf{(e)-(f)} Cleaning sample with water and IPA.}
    \label{sifig:smple_prep}
\end{figure}

\newpage

\subsection{Experimental displacement cross section}

\subsubsection{Dose until first defect}

The electron dose $\phi$ is defined as the number of the electrons hitting the specimen. 
The dose rate $\phi_R$ (s$^{-1}$) is defined by a beam current $I$ on the sample divided by the elementary charge
\begin{equation}
\phi_R = \frac{I}{e} = \frac{I_{\text{VOA}} \times C - I_{\text{offset}}}{e}.
\label{eq:phi_R}
\end{equation}

In the experimental setup the current can not be measured directly on the sample. From the metadata we know the virtual objective aperture (VOA) current $I_{\text{VOA}}$ and can calculate the current on the sample with the calibration constant $C$ and dark-current $I_\text{offset}$.

The time $t_d$ until the first defect occurred is the total number of pixels multiplied by the dwell time $t_{\text{dwell}}$
\begin{equation}
t_d = P \times t_{\text{dwell}}.
\label{eq:t_d}
\end{equation}

With equations \ref{eq:phi_R} and \ref{eq:t_d} we get the dose $\phi$ until the first defect 
\begin{equation}
\phi = \phi_R \times t_d.
\label{eq:phi}
\end{equation}

We assume that our experimental data follows a Poisson process and therefore the cumulative doses are exponentially distributed, obtained by cumulating them in discrete bins, where the first bin contains the total number of all events (counts) and in the next bin, only events with doses higher than the bin width remain. The same procedure is repeated for the following bins until the last one. Histograms containing the cumulative event counts with respect to the dose bins are shown in Supporting Fig.~\ref{sifig:expfit}.

This exponential fit-value is our expectation value $\lambda(E_\text{e})$ for the dose for each set of experimental parameters, most importantly the primary beam energy $E_\text{e}$. It is adapted from the probability density function 
\begin{equation}
f( \phi ) = \begin{cases}
\lambda(E_\text{e}) e^{-\lambda(E_\text{e}) \phi}  &  \text{for} \  x \ge 0 \\
0   &   \text{else}
\end{cases},
\label{eq:fphi}
\end{equation}
where $\phi$ is the irradiation dose.

Because both boron and nitrogen atoms can be displaced, but have different $T_d$, they must be analyzed separately. Correspondingly, the doses until the first defect need to be added up separately for boron and nitrogen. Since we want to measure the knock-on cross sections of the pristine material, defects with two or more atoms missing, which do happen often especially at higher primary beam energies (at 90~keV, this was the case in the majority of the experimental data series), need to be omitted from the analysis. We have also omitted cases where defects occur at the edge of the frame, because it is not clear whether those are single vacancies.

\begin{figure}
    \centering
    \includegraphics[width=17cm]{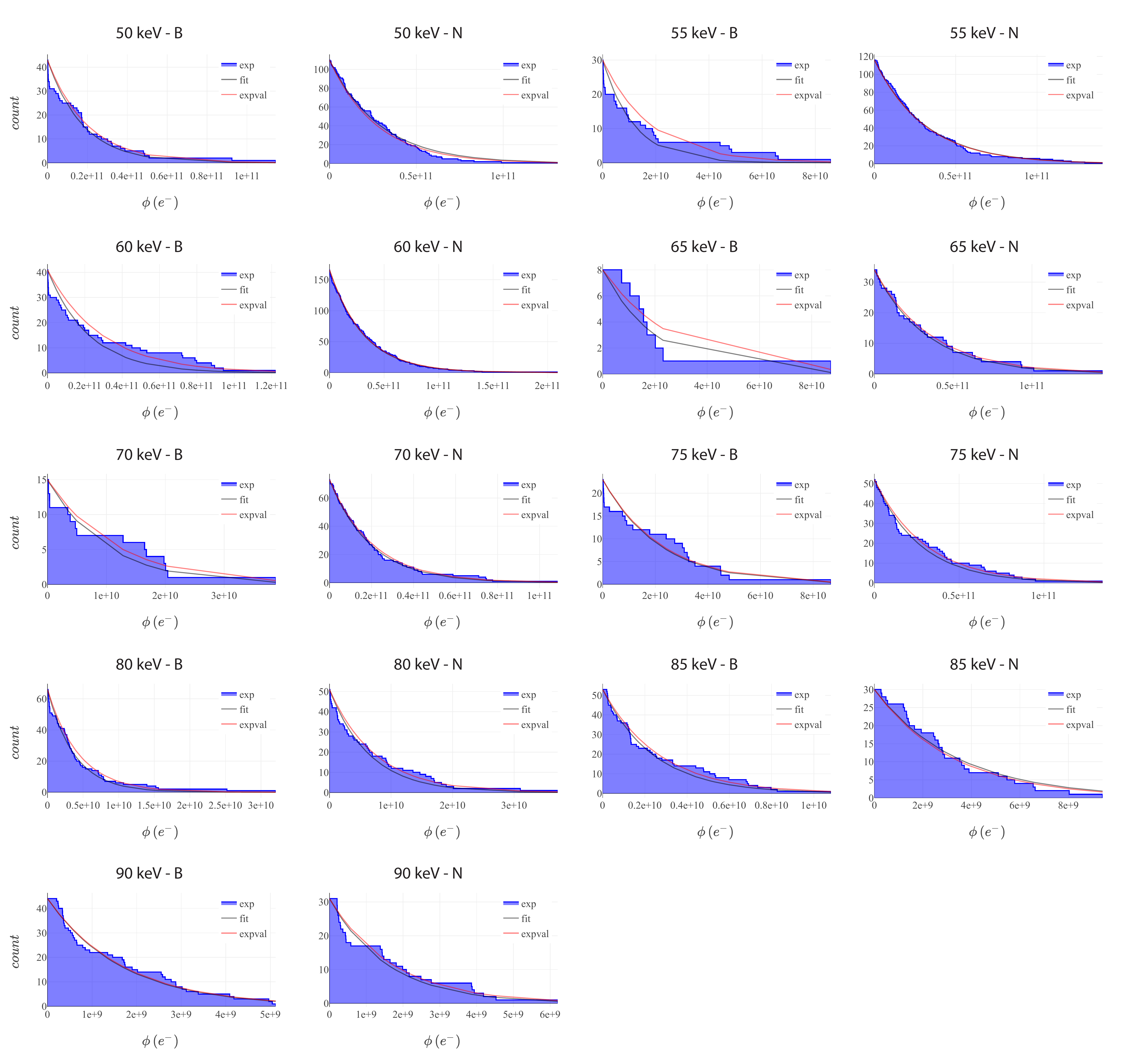}
    \caption{\textbf{Exponential fits of the Poisson expectation value for each element at each acceleration voltage.}}
    \label{sifig:expfit}
\end{figure}

\begin{figure}
    \centering
    \includegraphics[width=13cm]{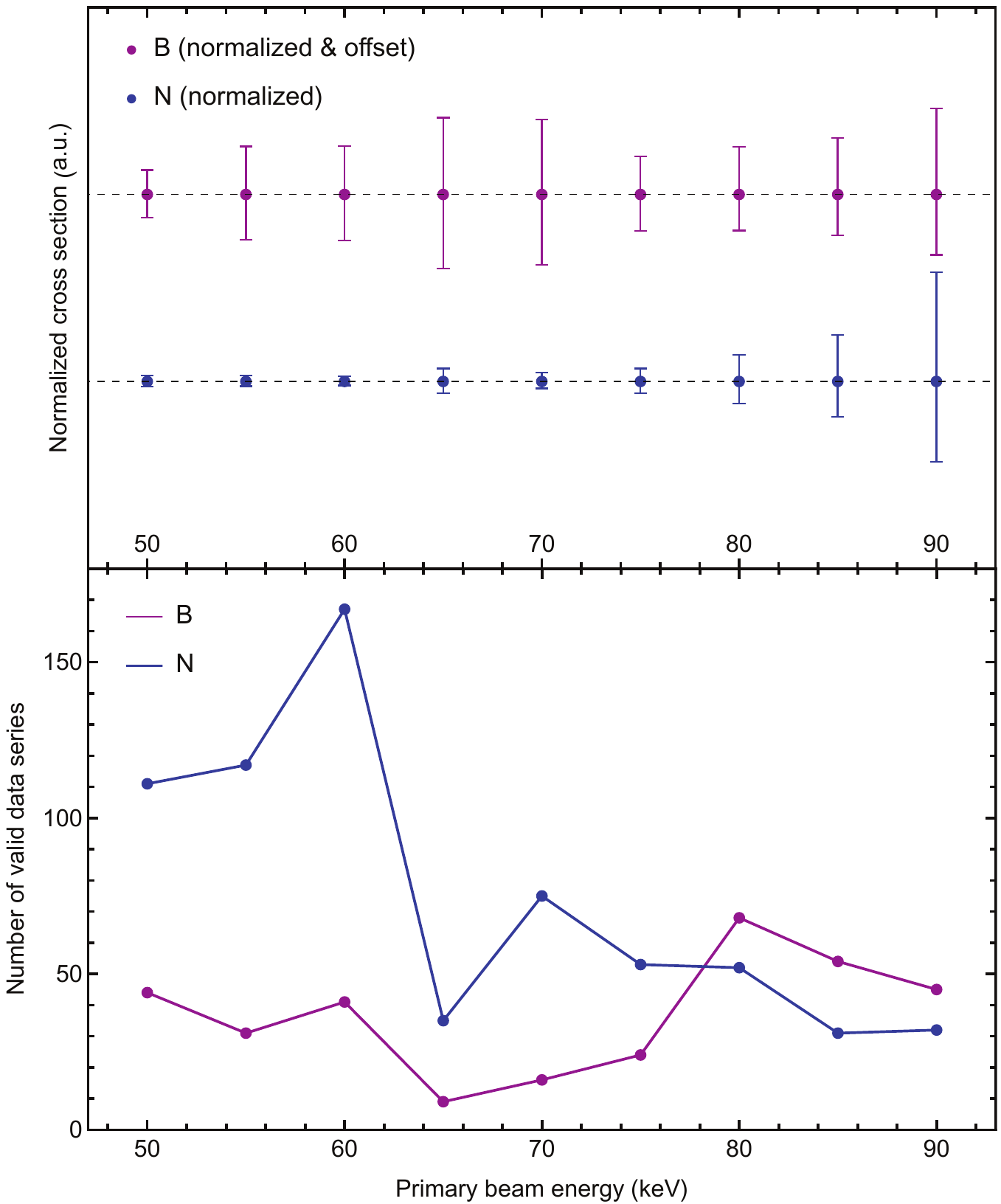}
    \caption{\textbf{Relative uncertainties of B and N.} (Top) Normalized relative uncertainties of B and N. (Bottom) Number of valid data series of each acceleration voltage.}
    \label{sifig:NB_errors_stat}
\end{figure}

\subsubsection{Experimental displacement cross sections}

The displacement cross sections $\sigma_\text{exp}$ of the experimental data at each primary beam energy $E_\text{e}$ were calculated via the expectation values $\lambda(E_\text{e})$ of the dose as described in the previous section and the areal density of hBN $\rho_{\text{hBN}}$ as
\begin{equation} 
\begin{split}
    \sigma_{\text{exp}} (E_\mathrm{e}) = \frac{1}{\rho_{\text{hBN}} \cdot \lambda(E_\text{e})}.
\end{split}
    \label{eq:sigma}
\end{equation}

The uncertainty of the experimental data was obtained by Gaussian error propagation, and the sum of uncertainty-weighted mean-squared errors was used for fitting the theoretical curves.

\begin{table}
\begin{tabular}{|c|c|c|c|c|c|c|} \hline
     & \multicolumn{3}{c|}{B} & \multicolumn{3}{c|}{N} \\ \hline
    $E$ (keV) & $\sigma$ (mb) & + $\Delta \sigma$ & - $\Delta \sigma$ & $\sigma$ (mb) & + $\Delta \sigma$ & - $\Delta \sigma$ \\ \hline
    50 & 28.5 & 4.0 & 3.8 & 17.2 & 0.5 & 0.5 \\ \hline
    55 & 47.7 & 11.1 & 10.5 & 15.9 & 0.5 & 0.4 \\ \hline
    60 & 30.4 & 6.0 & 5.7 & 18.4 & 0.5 & 0.4 \\ \hline
    65 & 34.3 & 10.9 & 10.6 & 25.5 & 1.8 & 1.6 \\ \hline
    70 & 63.3 & 18.0 & 17.0 & 32.1 & 1.5 & 1.2 \\ \hline
    75 & 26.1 & 5.0 & 4.8 & 21.2 & 1.5 & 1.3 \\ \hline
    80 & 142.1 & 24.1 & 18.2 & 83.8 & 12.0 & 9.9 \\ \hline
    85 & 235.1 & 68.5 & 49.5 & 161.7 & 40.1 & 30.8 \\ \hline
    90 & 321.5 & 149.8 & 106.5 & 357.7 & 209.0 & 153.6 \\ \hline
\end{tabular}
   \caption{Experimental cross sections $\sigma$ (in millibarn) for the displacement of boron and nitrogen from hBN as a function of the primary beam energy $E$, with corresponding asymmetric statistical uncertainties.}
    \label{Tab:cs_sigma}
\end{table}

\subsubsection{Machine-learning algorithm}
To ease a manual analysis of the images, a more accurate computer-assisted way was developed. A machine-learning algorithm not only detects the $x,y$ position of first occurring defect but also can tell by the intensity difference which element has been ejected. Supporting Figure \ref{sifig:ML} shows an example of one recorded series with 512$\times$512 pixels. The image shows the second frame, below which is a plot of the series analysis where detected defects are indicated by blue bars. The green line indicates where the first defect has been detected, which is the second frame in this case. With the integrated Gaussian filter applied to the image here it is easier to identify the different atoms by their contrast, which makes it possible to cross check the result manually.

\begin{figure}
    \centering
    \includegraphics[width=9cm]{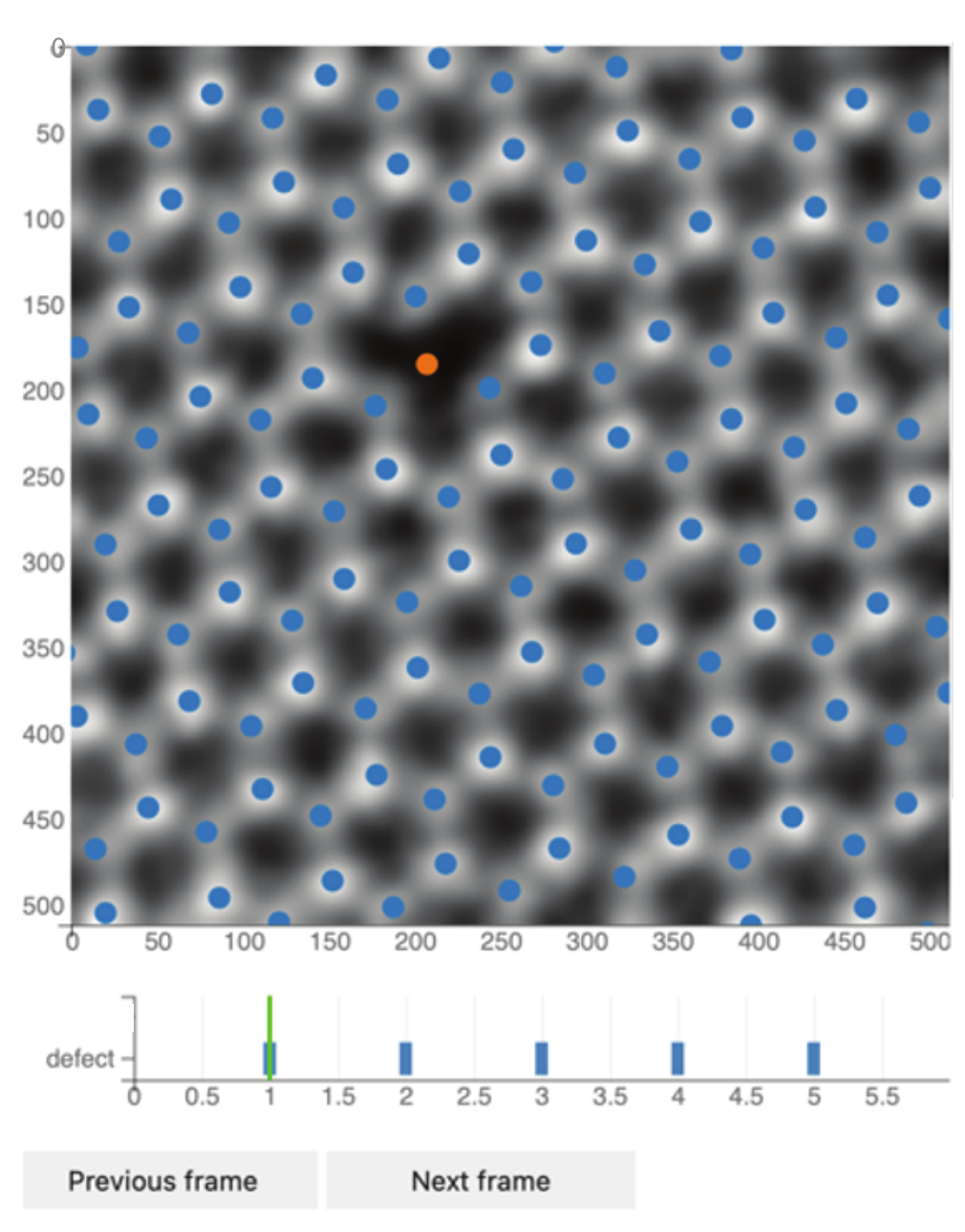}
    \caption{\textbf{Example result from the machine-learning algorithm.} The orange dot identifies the element (B or N) and the position of the of the defect. The plot below indicates the numbers of the frames with defects (frame two is shown).}
    \label{sifig:ML}
\end{figure}

\clearpage
\section{Ionization cross section}

To derive a practical expression for the probability of impact ionization, we start from the relativistic Bethe inelastic scattering cross section for electron shell $s$ from Williams \& Carter (Eq. 4.2 in Ref.~\citenum{williams_transmission_2009}):
\begin{equation}
\sigma_s=\left(\frac{\pi \mathrm{e}^{4} b_{\mathrm{s}} n_{\mathrm{s}}}{\left(\frac{\mathrm{m}_{0} v^{2}}{2}\right) E_{\mathrm{c}}}\right)\left[\log \left[c_{\mathrm{s}}\left(\frac{\mathrm{m}_{0} v^{2}}{2 E_{\mathrm{c}}}\right)\right]-\log \left(1-\beta^{2}\right)-\beta^{2}\right],\label{eq:WC}
\end{equation}
where $n_s$ is the number of electrons in the ionized subshell, $b_s$ and $c_s$, are constants for that shell, and $E_{c}$ is its critical ionization energy. Notably $m_0$ already is the relativistic rest mass (equal to 511 keV/c$^2$).

Since valence ionization energies are much smaller than those of core electrons, one could expect that they dominate the cross section, at least at relatively modest energies. If that assumption holds, we can use the Bethe formula only for the outermost shell, whose occupation $n_s$ is just the valence of the element ($n_2 = 3$ and $5$ for B and N, respectively). Thus, we can simply use the shell-dependent Bethe equation with known ionization energies and shell occupations, use that to fit the atomic data to obtain the constants $b_s$ and $c_s$, and then use these to describe the inelastic cross section in fitting our hBN displacement cross-section data. We should also note that this approach can be extended to other materials, most notably sulfur in MoS$_2$ as was done by Speckmann et al. based on this work~\cite{speckmann_combined_2023}.

Since Eq.~\ref{eq:WC} is in CGS units, to convert to SI units we replace $e^4 \rightarrow e^4/(4\pi \epsilon_0)^2$. To further explicitly account for the relativistic velocity of the beam electrons, we replace $v \rightarrow \beta c$. Finally, as it turned out that we need to consider more than one ionized state to accurately fit even the atomic data, let alone the hBN displacement data. As an approximation, we thus take the critical ionization energy to be the $i$th ionization energy of the atom, $E^{\mathrm{ion}}_i$. Thus, our converted Bethe equation reads
\begin{align}
\sigma_\mathrm{Bethe}({E_e, s, i})&=
\left(\frac{\pi \mathrm{e}^{4} b_{\mathrm{s}} n_{\mathrm{s}}}{(4\pi \varepsilon_0)^2 \left(\frac{m_0 c^2 \beta^2}{2}\right) E^{\mathrm{ion}}_i}\right) 
\left[\log \left[c_s \left(\frac{m_0 c^2 \beta^2}{2 E^{\mathrm{ion}}_i}\right)\right] - \log\left(1-\beta^{2}\right)-\beta^{2}\right] \nonumber \\
&= \left(\frac{2 \pi \mathrm{e}^{4} b_{\mathrm{s}} n_{\mathrm{s}}}{(4\pi \varepsilon_0)^2 m_0 c^2 \beta^2 E^{\mathrm{ion}}_i}\right) 
\left[\log \left[c_s \left(\frac{m_0 c^2 \beta^2}{2 E^{\mathrm{ion}}_i \left(1-\beta^{2}\right)}\right)\right]-\beta^{2}\right],\label{eq:bethe}
\end{align}
where the Lorentz factor $\beta$ can be expressed in terms of the primary beam energy $E_e$ as
\begin{equation}
  \beta = \beta(E_e) =
\sqrt{1-\left(1+\frac{E_{\mathrm{e}}}{m_{0} c^{2}}\right)^{-2}}.
\end{equation}

For boron, two ionized states give a decent description of the atomic data, though we have omitted the first five points from the fitting to ensure the higher-energy tail of the curve that is most relevant for our study is accurately fitted (Fig.~\ref{sifig_boron}). For nitrogen, three ionized states provide a very good fit without needing to drop any points (Fig.~\ref{sifig_nitrogen}), presumably due to its higher first ionization energy and the inclusion of a third state.

To estimate the uncertainty introduced by this fitting, we varied the number of omitted points for fitting with a single ionized state, and calculated the variation of the inelastic cross section at an experimental energy of 60 keV. We find that the variation was (2.95$\pm$0.12)$\times 10^{6}$ barn (4\%) for boron and (2.02$\pm$0.22)$\times 10^{6}$ barn (11\%) for nitrogen, and thus relatively negligible compared to other sources of uncertainty in our analysis.

\clearpage
\begin{figure}[h!]
  \includegraphics[width=0.85\textwidth]{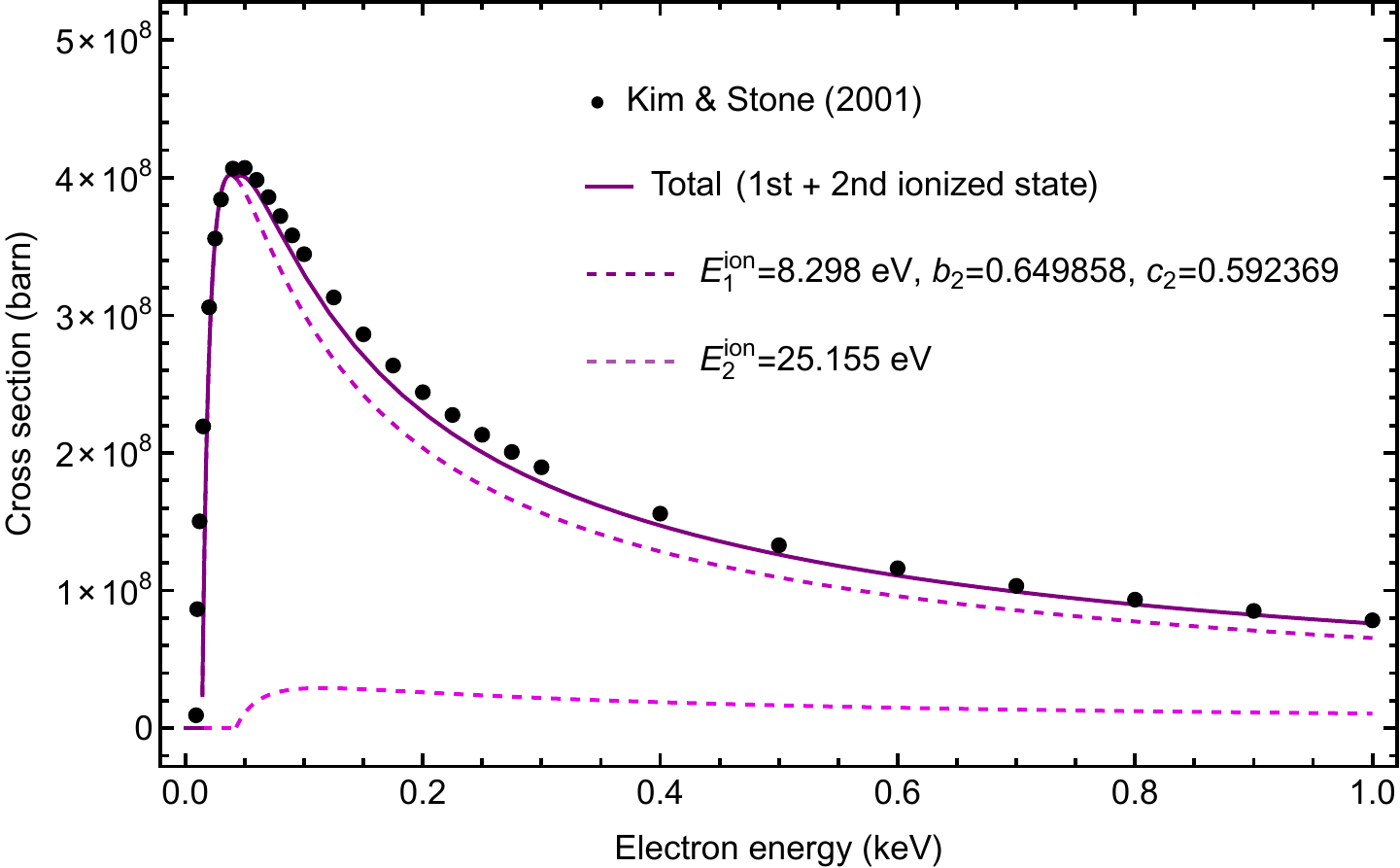}
  \caption{\textbf{Bethe fit for atomic cross section data for boron} ("Total counting" values from Table II of Kim \& Stone (2001), Ref.~\citenum{kim_ionization_2001}).}\label{sifig_boron}
\end{figure}

\begin{figure}[h!]
  \includegraphics[width=0.85\textwidth]{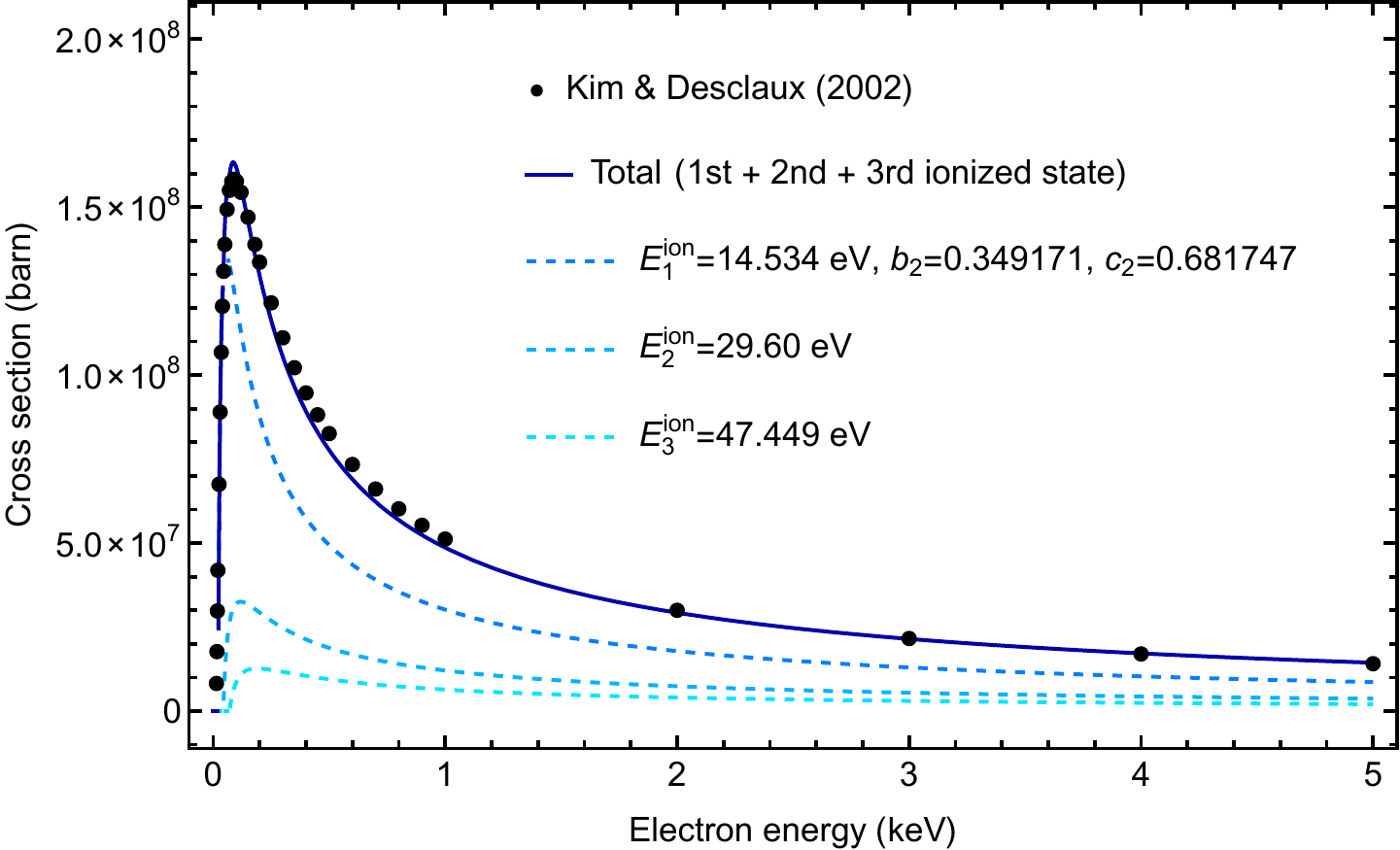}
  \caption{\textbf{Bethe fit for atomic cross section data for nitrogen} ("70-30 mix" values from Table IV of Kim \& Declaux (2002), Ref.~\citenum{kim_ionization_2002}).}\label{sifig_nitrogen}
\end{figure}

\begin{figure}[t]
    \centering
    \includegraphics[width=0.85\textwidth]{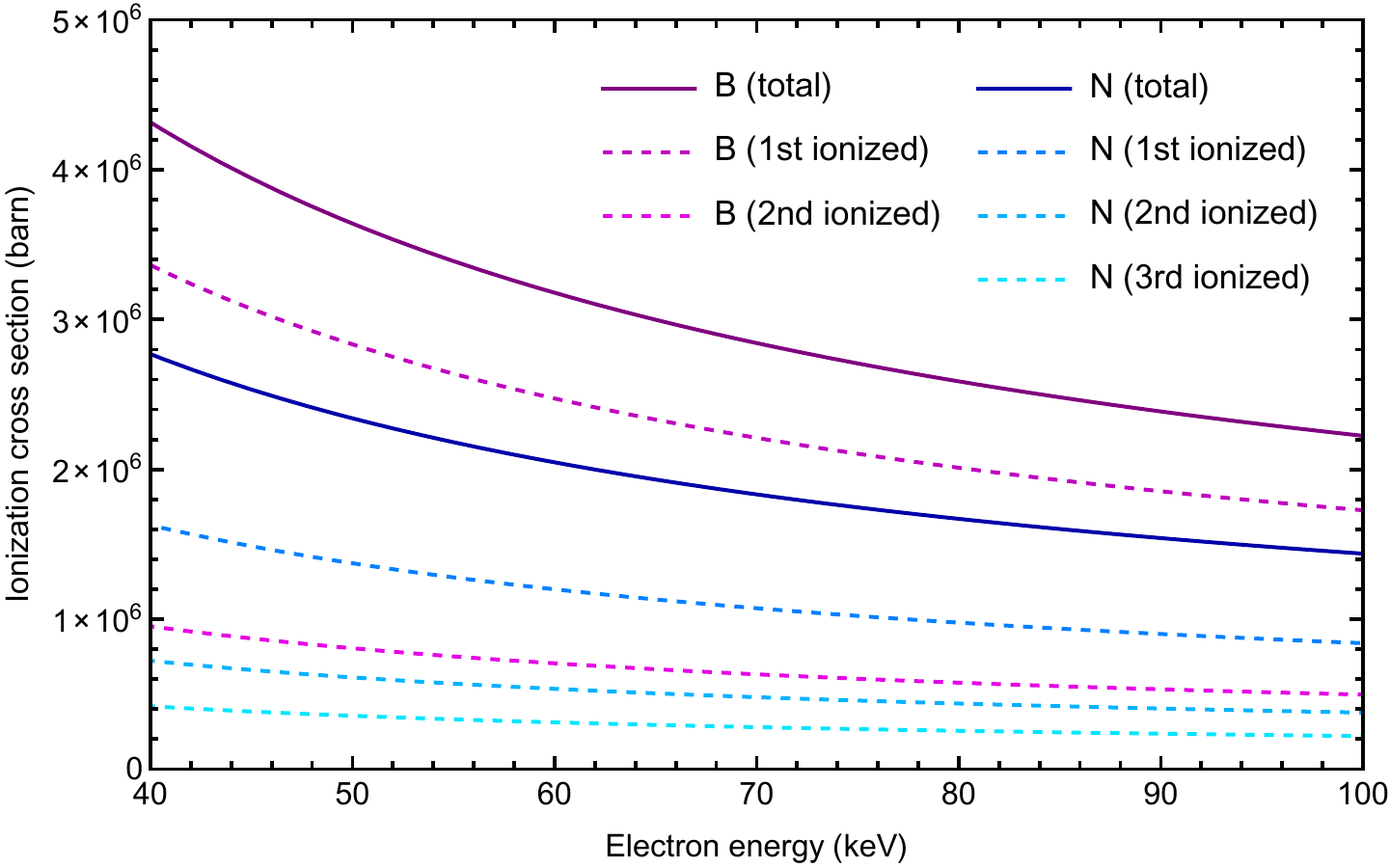}
    \caption{\textbf{Extrapolated theoretical valence ionization cross sections for B and N.} Bethe ionization cross sections for boron and nitrogen fitted to atomic scattering data shown in Supporting Figs.~\ref{sifig_boron} and~\ref{sifig_nitrogen} extrapolated to higher primary beam energies.}
    \label{sifig:ionization_B_N}
\end{figure}

\begin{figure}[t]
    \centering
    \includegraphics[width=0.75\textwidth]{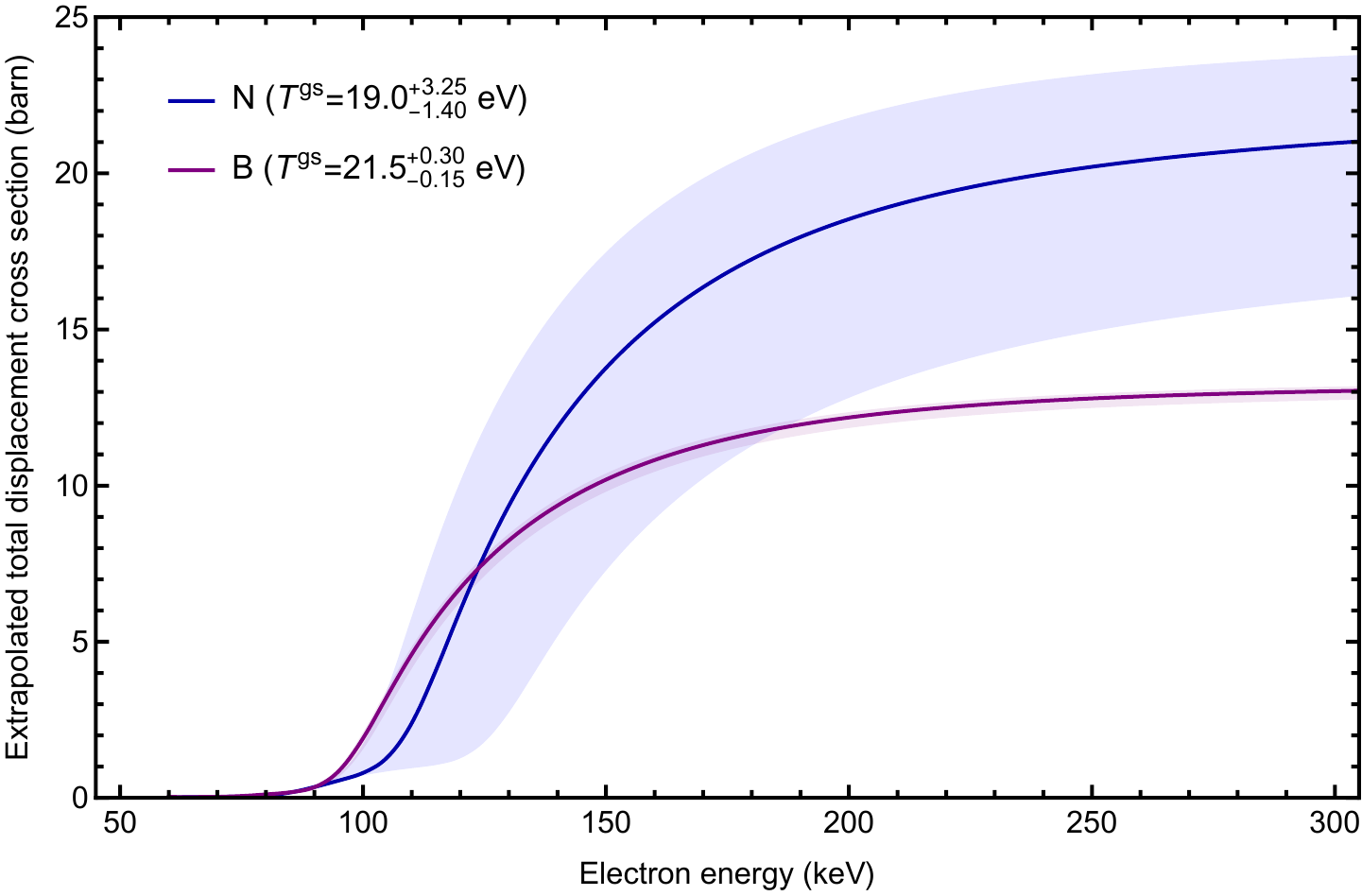}
    \caption{\textbf{Extrapolated total displacement cross sections for hBN.} Total cross sections for N and B displacement from hBN extrapolated from the fits in Figs.~\ref{fig:fit_B} and~\ref{fig:fit_N} to higher primary beam energies. The shaded areas correspond to uncertainties due to variation of threshold energies T$^\text{gs}$ where the uncertainty-weighted mean-squared sum of errors between the experimental data and the best-fit values are within $\pm$25\% (except for the upper limit for N, which corresponds to the DFT value of 22.25~eV).}
    \label{sifig:higherE}
\end{figure}

\begin{figure}[t]
    \centering
    \includegraphics[width=0.75\textwidth]{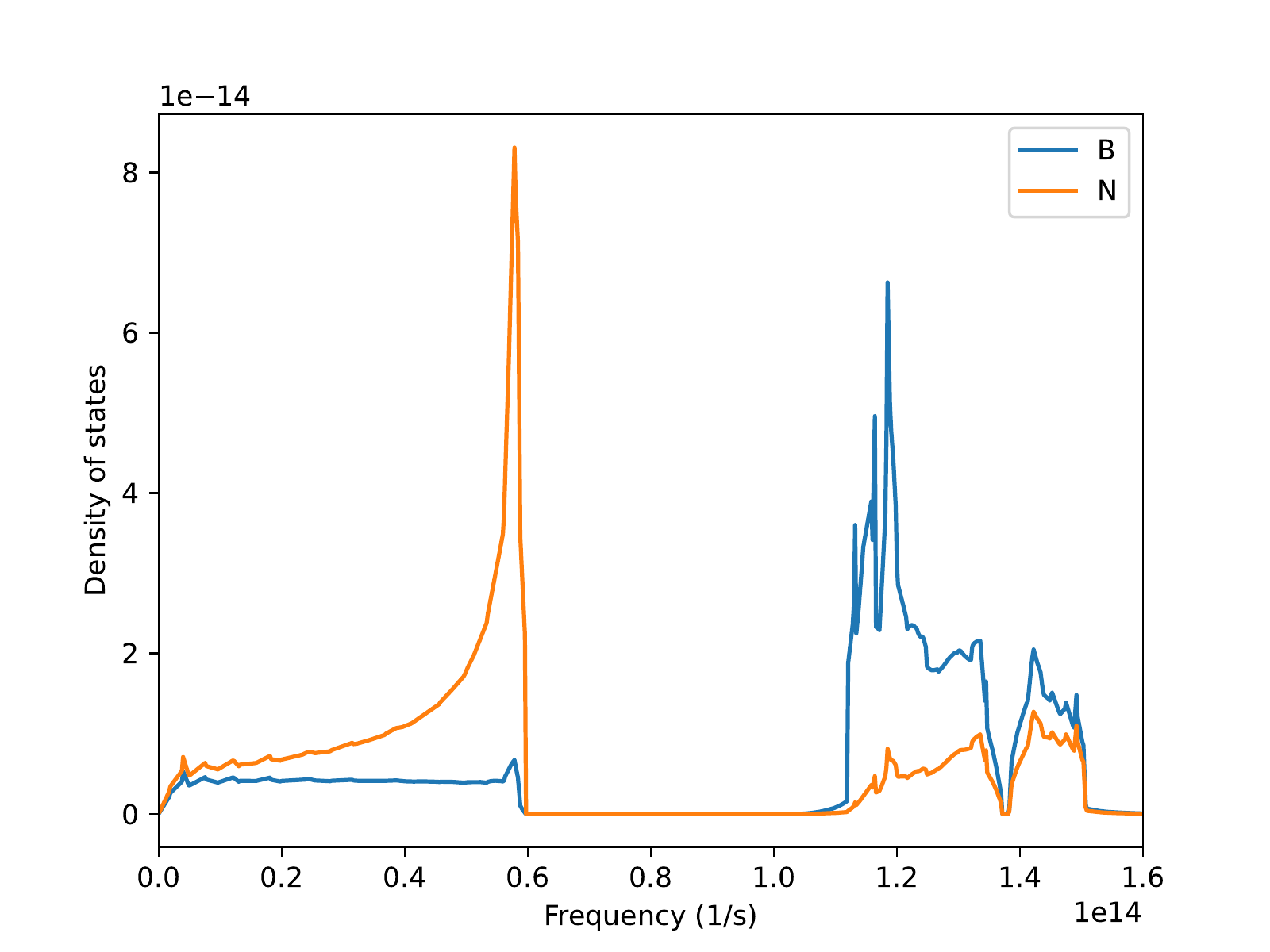}
    \caption{\textbf{Phonon density of states for hBN.} Element-decomposed phonon density of states calculated for hBN using density functional perturbation theory. The numerical data can be found as Supporting Data.}
    \label{sifig:phononDOS}
\end{figure}

\end{suppinfo}
\clearpage
\providecommand{\latin}[1]{#1}
\makeatletter
\providecommand{\doi}
  {\begingroup\let\do\@makeother\dospecials
  \catcode`\{=1 \catcode`\}=2 \doi@aux}
\providecommand{\doi@aux}[1]{\endgroup\texttt{#1}}
\makeatother
\providecommand*\mcitethebibliography{\thebibliography}
\csname @ifundefined\endcsname{endmcitethebibliography}
  {\let\endmcitethebibliography\endthebibliography}{}

\end{document}